\documentstyle[epsfig, 12pt]{article}
\begin{document}
\baselineskip18pt
\def\bpi{$B\to\pi$~}
\def\pslash{\rlap{\hspace{0.02cm}/}{p}}
\def\beq{\begin{eqnarray}}
\def\eeq{\end  {eqnarray}}
\def\GeV{{\rm GeV} }
\def\non{\nonumber}
\def\dirac#1{#1\llap{/}}
\def\as{\alpha_s}
\def\pv#1{\vec{#1}_\perp}
\def\lqcd{\Lambda_{\rm QCD}}

\newcommand\epjc[3]{Eur.\ Phys.\ J.\ C {\bf #1}, #3 (#2)}
\newcommand\ijmpa[3]{Int.\ J.\ Mod.\ Phys.\ A {\bf #1} (#2) #3}
\newcommand\jhep[3]{J.\ High Ener.\ Phys.\ {\bf #1} (#2) #3}
\newcommand\npb[3]{Nucl.\ Phys.\ B {\bf #1} (#2) #3}
\newcommand\npps[3]{Nucl.\ Phys.\ B (Proc.\ Suppl.) {\bf #1} (#2) #3}
\newcommand\plb[3]{Phys.\ Lett.\ B {\bf #1} (#2) #3}
\newcommand\prd[3]{Phys.\ Rev.\ D {\bf #1} (#2) #3}
\newcommand\prep[3]{Phys.\ Rep.\ {\bf #1} (#2) #3}
\newcommand\prl[3]{Phys.\ Rev.\ Lett.\ {\bf #1} (#2) #3}
\newcommand\rmp[3]{Rev.\ Mod.\ Phys.\ {\bf #1} (#2) #3}
\newcommand\sjnp[3]{{Sov.\ J.\ Nucl.\ Phys.\ }{\bf#1} (#2) #3}
\newcommand\yf[3]{{Yad.\ Fiz.\ }{\bf#1} (#2) #3}
\newcommand\zpc[3]{Z.\ Phys.\ C {\bf #1} (#2) #3}
\newcommand{\hepph}[1]{{\tt hep-ph/#1}}
\newcommand{\heplat}[1]{{\tt hep-lat/#1}}
\newcommand{\hepex}[1]{{\tt hep-ex/#1}}

\begin{flushright} BIHEP-TH-2002-9 \end{flushright}

\begin{center}
{ \Large\bf ~~\\  ~~\\
The systematic study of $B\to \pi$ form factors \\
  in pQCD approach and its reliability }
\end{center}

\vspace{0.5cm}
\centerline{ Zheng-Tao Wei $\rm ^{a,b}$~ and ~
             Mao-Zhi Yang  $\rm ^{a,c}$
 \footnote{e-mail address: weizt@itp.ac.cn;~~
  yangmz@mail.ihep.ac.cn}}

\vspace{0.5cm}
\begin{center}
$\rm ^{a}$ CCAST(World Laboratory), P.O.Box $8730$, Beijing $100080$,
 China\\
\vspace{0.3cm}
$\rm ^{b}$ Institute of Theoretical Physics, P.O.Box $2735$, Beijing
 $100080$, China \\
\vspace{0.3cm}
$\rm ^{c}$ Institute of High Energy Physics, P.O.Box $918(4)$, Beijing
 $100039$, China
\end{center}

\vspace{1cm}
\begin{abstract}
\vspace{0.2cm}\noindent The study of exclusive B decays in
perturbative QCD are complicated by the endpoint problem. In order
to perform the perturbative calculation, the Sudakov effects are
introduced to regulate the endpoint singularity. We provide a
systematic analysis with leading and next-to-leading twist
corrections for \bpi form factors in pQCD approach. The intrinsic
transverse momentum dependence of hadronic wave function and
threshold resummation effects are included in pQCD approach. There
are two leading twist B meson distribution amplitudes (or
generally wave functions) in general. The QCD equations of motion
provide important constraints on B meson wave functions. The
reliability of pQCD approach in \bpi form factors is discussed.
70\% of the result comes from the region $\alpha_s(t)/\pi<0.2$ and
38\% comes from the region where the momentum transfer $t\geq
1\GeV$. The conceptual problems of pQCD approach are discussed in
brief. Our conclusion is that pQCD approach in the present form
cannot provide a precise prediction for \bpi transition form
factors.
\end{abstract}

\baselineskip18pt
\newpage
\section{ Introduction }
The exclusive B decays provide an important test of the standard model of
particle physics. With the running of BaBar and Belle B-factories and the
proceeding of future B physics projects ( BTeV and LHCb etc.), a large
amount of B mesons will be accumulated to explore the origin of CP
violation and determine the CKM parameters, such as the angles $\alpha,
\beta$ and $\gamma$ in unitary triangle. In most cases, the limitation of
our theoretical ability prevents the precise prediction for exclusive B
decays so we have to refer to some phenomenological approaches.

The large B meson mass $m_B$ establishes a large scale so that
perturbative QCD (pQCD) may be applicable in exclusive B decays.
Recently, two different approaches based on perturbative QCD were
proposed to calculate the exclusive B decays. One approach,
usually say QCD factorization approach, states that two body
hadronic B decays can factorize and the relevant amplitude can be
written as the convolutions of non-perturbative quantities (the
light-cone distribution amplitudes of mesons and semi-leptonic
form factors) and perturbatively calculable hard scattering
kernels \cite{BBNS1}. The other approach is the modified pQCD, or
say pQCD approach for simplicity. In pQCD approach, the
semi-leptonic form factors for $B\to P$ transition (where P
represents light meson) are claimed to be perturbatively
calculable \cite{pQCD}.

The central difference between these two approaches is whether $B\to P$
transition form factors are perturbatively calculable, or specifically, whether
Sudakov effects can cure endpoint singularity. In $B\to P$ transition form
factors, the endpoint singularity is generated even at leading twist. In pQCD
approach, the transverse momentum are retained to regulate the endpoint
singularity and Sudakov double logarithm corrections are included to suppress
the long distance contributions from configuration of large transverse
separation. In \cite{Sachrajda}, the authors investigated the reliability of
Sudakov effects in \bpi form factors. Their conclusion is that Sudakov
suppression are so weak that they can not be applied in B decays. If their
criticism is right, it would bring the entire thought about Sudakov effects
into crisis.

In our previous study \cite{DHWY}, we investigated the Sudakov
effects in QCD factorization approach. Our conclusion is contrary
to the above criticism. Sudakov effects play an important role in
exclusive B decays. It is well known that the standard approach or
say BLER approach \cite{BLER} is questionable at experimentally
accessible energy scales, typically a few $\GeV$ region
\cite{Isgur}. This question also occurs in the hard spectator
scattering in $B\to P_1P_2$ decays. The QCD factorization approach
gives a finite result at leading twist. Carefully study shows that
there are large contributions coming from the region where the
momentum transfers are small. The introduction of Sudakov
suppression enlarges the application range of pQCD and makes it
self-consistent \cite{LiSterman}. Although the relevant numerical
result is power suppressed, the Sudakov effects is potentially
important. We disagree with the comment that the main purpose of
introduction of Sudakov suppression is to improve the computation
accuracy \cite{Sachrajda}. At twist-3 level, the chirally enhanced
power correction is logarithmically divergent, so the QCD
factorization approach can not apply. A phenomenological
parametrization method is designed to regulate the endpoint
singularity \cite{BBNS2}. This method is not self-consistent and
it introduces an arbitrary infrared cutoff. Sudakov suppression
establish a natural infrared cutoff. Obviously, pQCD approach has
the advantage of the parametrization method.

The disagreement of conclusions in \cite{pQCD, DHWY} and \cite{Sachrajda}
motivates us to reconsider the reliability of pQCD approach in \bpi form
factors. In this paper, we present a systematic study of \bpi form factors
in pQCD approach and examine the reliability of pQCD calculation. Compare with
the recent pQCD analysis of \bpi form factors \cite{Liform}, the present
analysis contains three new theoretical ingredients:
\begin{itemize}
 \item The intrinsic transverse momentum dependence of wave functions for B
  and pion mesons are included. The importance of intrinsic transverse
  momentum dependence is first pointed out in pion form factor in \cite{Jakob}.
  Because Sudakov suppression is not severely strong for real $m_B$, the effect
  of intrinsic transverse momentum dependence of wave function can not be
  neglected.
 \item B meson contains two leading wave functions. The assumption of single
  B meson wave function in the previous pQCD analysis is not valid. Equations
  of motion in HQET provide important constraint on the choice of B meson wave
  functions.
 \item The threshold resummation effect for B meson is taken into account.
  The perturbative analysis depends on the endpoint behavior of B meson
  distribution amplitudes. The jet function obtained from threshold resummation
  suppress the endpoint contribution, so it modifies the power behavior of
  endpoint contribution. We expect that it can improve the perturbative result
  since one of the two B meson distribution amplitudes does not vanish at
  endpoint.
\end{itemize}
In addition, the chirally-enhanced power corrections (twist-3 contribution)
are included in our analysis. It is noted that the contribution of
chirally-enhanced power corrections is comparable with or even larger than
leading twist contribution \cite{Liform}. Our study confirm it. The numerical
results of \bpi form factors at large recoil region in pQCD approach are
consistent with derivations of QCD sum rules \cite{Braun}.

The remainder of the paper is organized as follows. Sec. 2 introduces the
theoretical ingredients of pQCD approach. In sec. 3, we present the formulas
for \bpi form factors in pQCD approach, perform the numerical analysis,
and examine the reliability of pQCD approach. Finally, in sec. 4, the
conclusions and discussions are presented.

%%%%%%%%%%%%%%%%%%%%%%%%%%%%%%%%%%%%%%%%%%%%%%%%%%
%   The $B\to\pi $ form factors in pQCD approach
%%%%%%%%%%%%%%%%%%%%%%%%%%%%%%%%%%%%%%%%%%%%%%%%%%
\section { The $B\to\pi $ form factors in pQCD approach}

$B\to \pi$ form factors are the basic non-perturbative parameters in
semi-leptonic $B\to \pi l\nu$ decays and exclusive, nonleptonic decays
such as $B\to \pi\pi, \pi K$ decays in QCD  factorization approach. While
in pQCD approach, it can be perturbatively calculated from the universal
hadronic distribution amplitudes or wave functions. In this section, we
will give a general discussion of $B\to \pi$ form factors and introduce the
main ingredients of pQCD approach.

The form factors of $B\to \pi$ are defined by the following Lorentz
decompositions of biquark current matrix elements:
\beq \label{eq:bpi1}
 \langle \pi(P_{\pi})|\bar u \gamma_{\mu}b|\bar B(P_B)\rangle=
 (P_B+P_{\pi}-\frac{m_B^2-m_{\pi}^2}{q^2}q)_{\mu}F_+^{B\pi}(q^2)+
 \frac{m_B^2-m_{\pi}^2}{q^2} q_{\mu}F_0^{B\pi}(q^2) ,
\eeq
where $q=P_B-P_{\pi}$ is the momentum carried by lepton pair in semi-leptonic
$B\to \pi l\nu$ decays. At large recoil limit, $q^2=0$, the two form factors
$F_+^{B\pi}, F_0^{B\pi}$ reduce to one parameter, $F_+^{B\pi}(0)=F_0^{B\pi}(0)$.

First, we give our conventions on kinematics. We work in the rest frame of B
meson. The mass difference of b quark and B meson is negligible in the heavy
quark limit and we take approximation $m_b=m_B$ in our calculation. The masses
of light quarks $u$, $d$ and $\pi$ meson are also neglected. For discussion,
the momentum is described in terms of light-cone variables. We take
$k=(\frac{k^+}{\sqrt 2}, \frac{k^-}{\sqrt 2}, \vec k_{\bot})$ with
$k^{\pm}=k^0\pm k^3$ and $\vec k_{\bot}=(k^1, k^2)$. The scalar product of two
arbitrary vectors $A$ and $B$ is $A\cdot B=A_{\mu}B^{\mu}=
\frac{A^+B^- + A^-B^+}{ 2} - \vec A_{\bot}\cdot \vec B_{\bot}$.
The momentum of pion is chosen to be in the $``-"$ direction. Under these
conventions, $P_B    =(\frac{m_B}{\sqrt 2}, \frac{m_B}{\sqrt 2},\vec 0_\bot)$,
$P_{\pi}=(0, \eta \frac{m_B}{\sqrt 2},\vec 0_\bot)$,
$\bar P_{\pi}=(\eta \frac{m_B}{\sqrt 2}, 0,\vec 0_\bot)$ with
$\bar \eta=1-\eta=\frac{q^2}{m_B^2}$.
We define two light-like vectors $n_+\equiv(\sqrt 2,0,\vec 0_\bot)$ and
$n_+\equiv(0, \sqrt 2, \vec 0_\bot)$.

We denote $\xi$ as the momentum fraction of spectator anti-quark
in B meson, and $x$ as the  momentum fraction of the anti-quark in
pion. As plotted in Fig .\ref{figbpi}.

\subsection{ The physical picture of factorization in $B\to \pi$
form factors}

The most important ingredient of pQCD is factorization, i.e., the
separation of the long-distance dynamics from the short distance
dynamics which is pertubatively calculable. Although rigorous
proof of the factorization theorem is technically intricate, the
physical picture is simple and intuitive.

First, we illustrate the factorization of pion electromagnetic
form factor in order to introduce the basic idea of standard
approach and the modified approach. A highly virtual photon
collides on a quark in initial pion and makes it change its
direction. In exclusive process $\gamma^*\pi\to\pi$, the simple
power counting shows that at large momentum transfers, the valence
quarks dominate the process and the higher Fock states and the
intrinsic transverse momentum is power suppressed. The large
momentum $Q$ means the high resolution at the distance of $1/Q$.
In a small distance region, a parton only sees the other parton
relative to it and the initial and final pion can be considered as
a small-size color-singlet object during the hard interaction.
What is important for small-size color-singlet pion is that soft
gluon corrections cancel at the leading twist. This is the
well-known phenomena, ``color transparency''. The long-distance
interactions only occur before and  after the hard interaction, so
their effects can be factorized into the initial and final pions
respectively. Since the hard interaction is restricted in the
short distance, we only need the probability for the $q\bar q$
pair in pion to be within the transverse distance of $1/Q$, in
other words, the distribution amplitude of pion $\phi(x, Q^2)$.
The scale $Q$ acts as factorization scale as well as
renormalization scale. The above discussions can be grouped into
the standard factorization formula for electromagnetic form factor
of pion \beq
  F_{\pi}=\int dx dy \phi(x, Q^2) T(x, y, Q) \phi(y, Q^2).
\eeq

The standard factorization is proved successfully in the
asymptotic limit. The application of this standard approach at the
experimentally accessible energy scales, typically a few $GeV$
region, is criticized in \cite{Isgur}. The authors pointed out
that the perturbative calculations for electromagnetic form factor
have a large amount of contributions coming from soft endpoint
region ($x,y\to 0$) where the perturbative analysis is invalid.
This is the well-known ``endpoint" problem even there is no
endpoint singularity in the convolution of total amplitude. The
Sudakov effects are introduced to modify the endpoint behavior and
make pQCD applicable in few GeV region \cite{LiSterman}. The basic
idea of this modified approach, or say pQCD approach is using the
mechanism of Sudakov suppression to suppress long-distance
contributions of large transverse separations. Sudakov suppression
establishes a factorization scale $1/b$ in addition to the
momentum scale $Q$. The pion with small transverse separation has
small color dipole moment so that QCD factorization is revised. In
pQCD approach, Sudakov suppression plays a crucial role.

Second, We discuss the factorization in $B\to\pi$ form factors in
the large $m_B$ limit. B meson is a heavy-light system. It has
large size in longitudinal as well as transverse directions. If
the soft B meson wave functions overlap with the wave functions of
final meson, the separation of long-distance and short distance
dynamics is impossible. From this point, we  conclude that $B\to
D^{(*)}$ from factors in heavy quark limit are soft dominant and
cannot be calculated by pQCD. For $B\to \pi$ form factors, the
situation is different. The pion carries the energy of
$\frac{m_B}{2}$ when $q^2=0$.  When the fractional momentum of the
antiquark in pion is far away from the endpoint region, i.e. $x\gg
0$, the pion is highly restricted in longitudinal direction
because of Lorentz contraction. The soft gluons that attach to
small size color-singlet pion decouple in the large $m_B$ limit.
So factorization is applicable. Simply to say, the soft spectator
anti-quark in B meson must undergo hard strong interaction to
change into the fast moving parton in pion.

The above perturbative picture is destroyed by the endpoint
contribution where anti-quark in pion carries the momentum of
order $\lqcd$. In the standard approach, $B\to \pi$ form factor is
logarithmically divergent at leading twist (twist-2 for our case).
The divergence becomes more serious for next-to-leading twist
(twist-3) contribution and the linear divergence occurs. The
origin of this divergence is that the distribution amplitude does
not provide enough suppression at endpoint. In order to apply the
perturbative picture, the soft endpoint contribution must be
suppressed. As we have discussed for pion electromagnetic  form
factor, Sudakov effects can provide such suppression. With the
help of Sudakov suppression, pQCD factorization can be applicable.
In pQCD approach, the $B\to \pi$ form factors can be generally
written as convolutions of wave functions and hard scattering
kernels \beq \label{eq:pQCD1}
  F_{+,0}^{B\pi}=\int d\xi~ dx~ d^2 \vec b_B~ d^2 \vec b_\pi~
          \Psi_B(\xi, b_{B}, \mu)~ \Psi_{\pi}(x, b_{\pi}, \mu)~
          T_{+,0}(\xi, x, b_{B}, b_{\pi}, \mu) .
\eeq

%%%%%%%%%%%%%%%%%%%%%%%%%%%%%%%%%%%%%%%%%%%%%%%%%%
%   Pion meson wave functions
%%%%%%%%%%%%%%%%%%%%%%%%%%%%%%%%%%%%%%%%%%%%%%%%%%

\subsection{ Pion wave functions }
In the standard approach, the transverse momentum is assumed power
suppressed and neglected. When the two valence quarks in pion
carry large longitudinal momentum, i.e., the momentum fractions
$x, \bar x\gg 0$, the transverse momentum is power suppressed
compared to longitudinal momentum. In the endpoint region, the
transverse momentum is important and cannot be neglected. In pQCD
approach, transverse momentum is retained in hard scattering
kernels and the non-perturbative parameters are hadronic wave
functions in general.

The two valence quarks in pion have transverse momentum as well as
the longitudinal momentum. Compared to the collinear limit, the
momentum of the quarks in pion ( with momentum $P_\pi$ ) changes
to \beq k_1=xp_\pi+k_{\bot}, ~~~~~~k_2=\bar x p_\pi-k_{\bot} ,
\eeq where $x$ and $\bar x$ denote the longitudinal momentum
fractions of quark and anti-quark, respectively. For our case, the
meson is on-shell and the partons are slightly off-shell. The
off-shellness of the parton is proportional to $k_{\bot}^2$ which
is power suppressed compared to $m_{B}^2$.

As we have discussed above, the intrinsic transverse momentum
dependence of wave functions of pion and B meson can not be
neglected because Sudakov suppression is not strong in the real
$m_B$ energy.  The distribution amplitude can be obtained from the
integration of wave function over the transverse momentum \beq
\label{eq:ds} \phi(x)=\int_{|k_{\bot}|<\mu} d^2k_{\bot} \Psi(x,
k_{\bot}), \eeq where $\mu$ is the ultraviolet cutoff.

Similar to the definition of distribution amplitudes with leading and
next-to-leading twist \cite{Brauntwist}, the light-cone wave functions of
pion are defined in terms of bilocal operator matrix element
\beq  \label{eq:pionwavefunctions}
 \langle \pi(p)|\bar q_\beta(z) q_\alpha|0\rangle
 &=& \frac{i f_{\pi}}{4} \int_0^1 dx\int d^2 \vec k_\bot
    e^{i(x p\cdot z- \vec k_\bot \cdot \vec z_\bot)}  \\
&\times& \left\{ \pslash \gamma_5 \Psi_{\pi}(x, k_\bot)
    - \mu_{\pi}\gamma_5 \left( \Psi_p(x, k_\bot)
    - \sigma_{\mu\nu} p^\mu z^\nu \frac{\Psi_\sigma(x,k_\bot)}{6}
    \right) \right\}_{\alpha\beta},  \nonumber
\eeq where $f_{\pi}$ is the decay constant of pion. The parameter
$\mu_{\pi}=m_{\pi}^2/(m_u+m_d)$ for charged pion. For neutral
pion, we use the same parameter $\mu_{\pi}$ as the charged pion.
$\Psi_{\pi}$, $\Psi_p$ and $\Psi_{\sigma}$ are the twist-2 and
twist-3 wave functions,
respectively. %In the asymptotic limit,
%$\phi_{\pi}(x)=6x\bar x$, $\phi_p(x)=x$ and $\phi_{\sigma}(x)=6u\bar x$.
The twist-3 wave functions contributes power corrections. But at
$m_B$ energy scale, the chirally enhanced parameter
$r_{\pi}=\frac{\mu_{\pi}}{m_B}\sim{\cal O}(1)$ is not small. So
the twist-3 contribution should be considered in B decays.

In order to perform the analysis in pQCD approach, we need to
transform the parameters in terms of coordinate variable in
Eq.(\ref{eq:pionwavefunctions}) into the momentum space
configuration. We use the momentum projection given in
\cite{BenekeFeldmann}: \beq \label{eq:pionprojector}
   M_{\alpha\beta}^{\pi} = \frac{i f_{\pi}}{4} \Bigg\{
   \pslash\,\gamma_5\,\Psi_{\pi} - \mu_P\gamma_5 \left(
   \Psi_p - i\sigma_{\mu\nu}\,\frac{p^\mu\bar p^\nu}{p\cdot\bar p}\,
   \frac{\Psi_{\sigma}'}{6}
   + i\sigma_{\mu\nu}\,p^\mu\,\frac{\Psi_\sigma}{6}\,
   \frac{\partial}{\partial k_{\perp\nu}} \right)
   \Bigg\}_{\alpha\beta},
\eeq where $\Psi_{\sigma}'=\frac{\partial \Psi_{\sigma} (x,
k_{\bot})}{\partial x}$. The wave functions $\Psi_{\pi},  \Psi_p,
\Psi_{\sigma}$ may have different transverse momentum dependence,
this will make the calculation difficult. In order to simplify
discussions, we assume the same transverse momentum dependence for
these wave functions.

In pQCD approach, the convolutions of wave functions and hard scattering
kernel are presented in transverse configuration $b$-space. We need to
define wave function in $b$-space through Fourier transformation
\beq  \label{eq:wavefunctionb}
\Psi(x, b)=\int d^2{\vec k_{\bot}}~e^{-i\vec k_{\bot}
             \cdot \vec b}\Psi(x, k_{\bot}).
\eeq %
From Eq.(\ref{eq:ds}) and (\ref{eq:wavefunctionb}), we can obtain%
\beq \label{eq:zero} \phi(x)=\Psi(x, b=0). \eeq%
The impact parameter $b$ represents the transverse separation of
quark and anti-quark in pion. Eq.(\ref{eq:zero}) shows that the
distribution amplitude is equal to the wave function at zero
transverse separation.

The intrinsic transverse momentum dependence of wave function of
pion is unknown from the first principle in QCD. We take a simple
model in which the dependence of the wave function on the
longitudinal and transverse momentum can be separated into two
parts:%
\beq \Psi(x, k_{\bot})=\phi(x) \times \Sigma(k_{\bot}) ,
\eeq %
where $\phi(x)$ is the pion distribution amplitude. $\phi(x)$ and
$\Sigma(k_{\bot})$ satisfy the normalization conditions%
\beq \int_0^1\phi(x)=1,~~~~~~~~ \int d^2 \vec k_{\bot}
\Sigma(k_{\bot})=1. \eeq%
The $k_{\bot}$ dependence of the wave function is contained in
$\Sigma(k_{\bot})$. In \cite{Jakob}, $\Sigma(k_{\bot})$ is assumed
to be a Gaussian distribution, %
\beq
\Sigma(k_{\bot})=\frac{\beta^2}{\pi}\exp(-\beta^2 k_{\bot}^2),
\eeq %
Transforming it into the transverse configuration b-space,  we
obatin \beq \Sigma(b)=\int d^2{\vec k_{\bot}}~e^{-i\vec k_{\bot}
             \cdot \vec b}\Sigma(k_{\bot})
            =\exp(-\frac{b^2}{4\beta^2}),
\eeq The oscillating parameter $\beta$ is fixed by requiring the
root mean square transverse momentum (r.m.s.), $\langle k_{\bot}^2
\rangle^{1/2}$ is the order of $\Lambda_{QCD}$. Their relation can
be obtained from %
\beq \langle k^2_\bot\ \rangle=\frac{\int_0^1 dx
\int d^2 \vec k_\bot
  k_\bot^2|\Psi(x,k_\bot)|^2}
 {\int_0^1 dx \int d^2 \vec k_\bot
 |\Psi(x,k_\bot)|^2}=\frac{1}{2\beta^2}.
\eeq%
Thus, $\langle k^2_\bot \rangle^{1/2}=\frac{1}{\sqrt 2 \beta}$. If
the root mean square transverse momentum $\langle k^2_\bot
\rangle^{1/2}=0.35\GeV$, then $\beta^2=4\GeV^{-2}$. The detailed
discussions about choice of the oscillation parameter $\beta$ can
be found in \cite{Jakob}.

%%%%%%%%%%%%%%%%%%%%%%%%%%%%%%%%%%%%%%%%%%%%%%%%%%
%   B meson wave functions
%%%%%%%%%%%%%%%%%%%%%%%%%%%%%%%%%%%%%%%%%%%%%%%%%%
\subsection{ B meson wave functions }

The intrinsic dynamic of B meson is different from the case of
light pion meson. The momentum components of the spectator quark
$l$ are of order $\Lambda_{QCD}$. The most convenient tool to
describe B meson is heavy quark effective theory (HQET). Since we
have chosen $P_{\pi}$ in the $``-"$ direction, the hard scattering
amplitude does not dependent on $l_-$, the $l_-$ dependence of the
wave functions can be integrated out. In HQET, the B meson wave
functions are defined by the general Lorentz decomposition of the
light-cone matrix element \cite{GrozinNeubert, Sachrajda}%
\beq
\langle 0|\bar{q}_\beta(z) b_\alpha(0)|\bar{B}(p_{B})\rangle
  =-\frac{if_B}{4} \left\{ \frac{\dirac{p}_B+m_B}{2}
    \left[ 2\tilde\Psi^+_B(z^2, t)
    +\frac{\tilde\Psi^-_B(z^2, t)-\tilde\Psi^+_B(z^2, t)}{t}\dirac{z}
    \right] \gamma_5 \right\} _{\alpha\beta},
\eeq
with $v=\frac{p_B}{m_B}$ and $t=v\cdot z$. A path-ordered exponential
is implicitly present in the gauge-independent matrix element.

The wave functions in momentum space can be obtained through the Fourier
transformation:
\beq
\Psi_B^\pm(l_+, l_{\bot})=
 \frac{1}{2} \int \frac{d^2\pv{z}}{(2\pi)^2} \frac{dz_-}{2\pi}
    e^{i(l_+z_-/2-\pv{l}\cdot\pv{z})} \tilde\Psi^\pm_B(-{\pv{z}\,}^2,z_-/2)\
    ,
\eeq
where $l=(l_+/\sqrt{2},0,\pv{l})$ and $z=(0,z_-/\sqrt{2},\pv{z})$,
so that $z^2=-\pv{z}^{\,2}$ and $t=z_-/2$.

A momentum projection for the matrix element of B meson is needed
to simplify the calculation. In \cite{BenekeFeldmann}, the authors
obtain a projection which is valid for distribution amplitudes. We
extend their analysis to include the transverse momentum
dependence of wave functions. We observe that their formulas need
no modification except the replacement of distribution amplitudes
by wave functions. The proof of this point is given in Appendix.
So, the momentum-space projection operator for B meson is:
\beq
M^B_{\alpha\beta}=-\frac{if_B}{4}
  \left\{\frac{\dirac{p}_B+m_B}{2}\left[
  \Psi^+_B \dirac{n}_+ + \Psi^-_B \dirac{n}_-
  -\Delta(l_+, l_{\bot}) \gamma^\mu \frac{\partial}
  {\partial l_\perp^\mu}
  \right]\gamma_5\right\}_{\alpha\beta}\ , \label{eq:bmesonproj}
\eeq
where $\Delta(l_+, l_{\bot})=\int_0^{l^+} dl
(\Psi^-_B(l, l_{\bot})-\Psi^+_B(l, l_{\bot}))$.

The projection operator can be represented in the form which is helpful to
compare with the results in the previous pQCD analysis
\cite{Liform}
\beq
M^B_{\alpha\beta}=-\frac{if_B}{4}
  \left\{(\dirac{p}_B+m_B)\left[
  \Psi_B  + \frac{\dirac{n}_+-\dirac{n}_-}{2}\bar\Psi_B
  -\frac{1}{2}\Delta(l_+, l_{\bot}) \gamma^\mu \frac{\partial}
  {\partial l_\perp^\mu}
  \right]\gamma_5\right\}_{\alpha\beta},
\eeq
where $\Psi_B, \bar\Psi_B$ are defined by
\beq
    \Psi_B=\frac{\Psi_B^+ + \Psi_B^-}{2}, ~~~~~
\bar\Psi_B=\frac{\Psi_B^+ - \Psi_B^-}{2}.
\eeq

In \cite{BBNS1}, the authors consider a different prjection
operator for B meson
\beq \label{eq:BBNS2} %
M^B_{\alpha\beta}=-\frac{if_B}{4}
  \left\{(\dirac{p}_B+m_B)\gamma_5 [
  \Psi_{B1}  + \dirac{n}_-\Psi_{B2} ]
  \right\}_{\alpha\beta}, %
\eeq %
where $\Psi_{B1}, \Psi_{B2}$ are defined by %
\beq %
\Psi_{B1}=\Psi_B^+, ~~~~~ \Psi_{B2}=\frac{\Psi_B^+ - \Psi_B^-}{2}.
\eeq

The above three projection operators are equivalent up to leading
power in $1/m_b$. The projection given in Eq.(\ref{eq:BBNS2})
neglects the $\Delta$ term which is proportional to $l^+/m_b$ and
thus power suppressed.

The B meson distribution amplitudes are obtained from transverse
momentum integral of the relevant wave functions or from wave
functions at zero transverse separation. $\phi_B^{\pm}$ and
$\phi_B, \bar\phi_B$ are distribution amplitudes relative to  wave
functions of $\Psi_B^{\pm}$ and $\Psi_B, \bar\Psi_B$ respectively
and they are related by %
\beq
    \phi_B=\frac{\phi_B^+ + \phi_B^-}{2}, ~~~~~
\bar\phi_B=\frac{\phi_B^+ - \phi_B^-}{2}
\eeq

The equations of motion impose constraint on the wave functions.
Using the equation of motion for the light quark of B meson and
neglecting the effects of three-parton and higher Fock states, we
can obtain \cite{BenekeFeldmann, Sachrajda}%
\beq
\left.\frac{\partial\tilde\Psi^-_B}{\partial t}
  +\frac{\tilde\Psi^-_B-\tilde\Psi^+_B}{t}
 \right|_{z^2=0} &=& 0\ ,  \label{e22}\\
\left.\frac{\partial\tilde\Psi^+_B}{\partial z^2}
     +\frac{1}{4}\frac{\partial^2\tilde\Psi^-_B}{\partial t^2}
 \right|_{z^2=0} &=& 0\ . \label{e23}
\eeq

Similar to the discussion for pion wave function, we consider a model in
which the dependence on the longitudinal and transverse momenta of B meson
is factorized:
\beq\label{eq:separable}
\Psi_B^\pm(\xi, l_{\bot})=\phi_B^\pm(\xi)\times \Sigma_B^\pm(l_\bot),
\eeq
where $\xi=\frac{l_+}{m_B}$  is the longitudinal momentum fraction.
The $\phi^{\pm}_B$ are the two distribution amplitudes of B mesons.
The normalization conditions are
\beq
\int d\xi \phi_B^\pm(\xi)=1, ~~~~~~~~~~~~
\int d^2\pv{l}\Sigma_B^\pm(l_\bot)=1.
\eeq

Under the above assumptions, the two constraints become %
\beq
\phi_B^+(\xi)&=&-\xi \frac{d\phi_B^-}{d\xi}(\xi),\\
\xi^2 m_B^2 \phi_B^-(\xi)&=&\omega_B^2 \phi_B^+(\xi), \eeq %
where
$\omega_B^2=\int d^2\pv{l} \pv{l}^2 \Sigma_B^+(l_\perp)$.

The distribution amplitudes of $\phi_B^\pm$ can  be solved
analytically and the corresponding solution is:
\beq \label{eq28}
\phi^-_B(\xi)=\sqrt\frac{2}{\pi}\frac{m_B}{\omega_B}
  \exp(-\frac{\xi^2 m_B^2}{2 \omega_B^2}), ~~~~~~~~~~~~
  \phi^+_B(\xi)=\sqrt{\frac{2}{\pi}}
  \frac{\xi^2 m_B^3}{\omega_B^3}
  \exp(-\frac{\xi^2 m_B^2}{2 \omega_B^2}).
\eeq

The parameter $\omega_B$ is fixed by the value of root mean square
transverse momentum of B meson. For the transverse momentum
dependent function, we choose the Gaussian distribution:
\begin{equation}
\Sigma_B^+(l_\perp)=\frac{1}{2\pi\omega'^{2}_B}
   \exp (-\frac{l_\perp^2}{2\omega'^{2}_B}).
\end{equation}
The parameters $\omega_B$ and $\omega'_B$ are not independent, they are
related by $\omega_B=\sqrt 2 \omega'_B$. The root mean square transverse
momentum $\langle l^2_\bot \rangle^{1/2}=\frac{\omega_B}{\sqrt 2}=\omega_B'$.
In transverse configuration b-space, the transverse momentum dependent
function is
\beq \label{eq:ktb}
 \Sigma_B^+(b_B)=\exp(-\frac{\omega^{'2}_B b_B^2}{2}),
\eeq
where $b_B$ is the conjugated variable of $\l_{\bot}$. For $\Sigma^-_B(b_B)$
function, we assume it is the same as $\Sigma^+_B(b_B)$.

A more rigorous study of B meson wave functions in HQET is given
recently in \cite{Qiao}. The authors find an analytic solution in
the heavy quark limit using the QCD equations of motion. The
transverse momentum distribution can also be derived without the
assumption of Gauss behavior. The obtained B meson wave functions
is model-independent.

%At last, we would like to denote the difference of the
%distribution amplitudes between the light pion and the heavy B
%meson. According to \cite{GrozinNeubert}, the B meson distribution
%can also be defined by the matrix elements of the pseudoscalar,
%axial, and tensor currents. For light pion meson, the matrix
%element of the axial current contributes the leading contribution
%and the pseudoscalar, tensor currents contributes the power
%corrections. While for the B meson, all the contributions are
%leading in the large $m_B$ limit. Only choose the matrix element
%of the axial current for B meson may lost large contribution.

%%%%%%%%%%%%%%%%%%%%%%%%%%%%%%%%%%%%%%%%%%%%%
%   Sudakov form factor
%%%%%%%%%%%%%%%%%%%%%%%%%%%%%%%%%%%%%%%%%%%%%
\subsection{ Sudakov form factor }

There are two types of resummation: Sudakov resummation (or say
$b$-space resummation) and threshold resummation. These two
resummation effects lead to suppression in different space: the
region with large transverse separations $b$ for Sudakov
resummation and the small longitudinal fractional momentum $x$
region for threshold resummation.

First, we discuss the Sudakov resummation in brief. At $\as$
order, the overlap of soft and collinear divergences produce
double logarithms $-c\ln^2 Qb$. The transverse impact parameter
$b$ is used to regulate the infrared divergence. In transverse
configuration $b$ space, the Sudakov double logarithms are resumed
up to next-to-leading-log approximation. A exponential factor
$e^{-s(x, b, Q)}$ will be obtained from $b$-space resummation. we
present the explicit expression of the exponent $s(x,b,Q)$
appearing in Sudakov form factor.

Define the variables,
\beq
{\hat q} \equiv  {\rm ln}\left(xQ/(\sqrt 2\Lambda_{QCD})\right),~~~~~~
{\hat b} \equiv  {\rm ln}(1/b\Lambda_{QCD}).
\eeq
According to \cite{LiSudakov}, the exponent $s(x,b,Q)$ is presented up to
next-to-leading-log approximation
\beq
&& s(x,b,Q)=
\nonumber \\
&&~~\frac{A^{(1)}}{2\beta_{1}}\hat{q}\ln\left(\frac{\hat{q}}
{\hat{b}}\right)-
\frac{A^{(1)}}{2\beta_{1}}\left(\hat{q}-\hat{b}\right)+
\frac{A^{(2)}}{4\beta_{1}^{2}}\left(\frac{\hat{q}}{\hat{b}}-1\right)
%\nonumber \\
-\left[\frac{A^{(2)}}{4\beta_{1}^{2}}-\frac{A^{(1)}}{4\beta_{1}}
\ln\left(\frac{e^{2\gamma_E-1}}{2}\right)\right]
\ln\left(\frac{\hat{q}}{\hat{b}}\right)
\nonumber \\
&&+\frac{A^{(1)}\beta_{2}}{4\beta_{1}^{3}}\hat{q}\left[
\frac{\ln(2\hat{q})+1}{\hat{q}}-\frac{\ln(2\hat{b})+1}{\hat{b}}\right]
+\frac{A^{(1)}\beta_{2}}{8\beta_{1}^{3}}\left[
\ln^{2}(2\hat{q})-\ln^{2}(2\hat{b})\right]
\nonumber \\
&&+\frac{A^{(1)}\beta_{2}}{8\beta_{1}^{3}}
\ln\left(\frac{e^{2\gamma_E-1}}{2}\right)\left[
\frac{\ln(2\hat{q})+1}{\hat{q}}-\frac{\ln(2\hat{b})+1}{\hat{b}}\right]
-\frac{A^{(2)}\beta_{2}}{16\beta_{1}^{4}}\left[
\frac{2\ln(2\hat{q})+3}{\hat{q}}-\frac{2\ln(2\hat{b})+3}{\hat{b}}\right]
\nonumber \\
& &-\frac{A^{(2)}\beta_{2}}{16\beta_{1}^{4}}
\frac{\hat{q}-\hat{b}}{\hat{b}^2}\left[2\ln(2\hat{b})+1\right]
+\frac{A^{(2)}\beta_{2}^2}{432\beta_{1}^{6}}
\frac{\hat{q}-\hat{b}}{\hat{b}^3}
\left[9\ln^2(2\hat{b})+6\ln(2\hat{b})+2\right]
\nonumber \\
&& +\frac{A^{(2)}\beta_{2}^2}{1728\beta_{1}^{6}}\left[
\frac{18\ln^2(2\hat{q})+30\ln(2\hat{q})+19}{\hat{q}^2}
-\frac{18\ln^2(2\hat{b})+30\ln(2\hat{b})+19}{\hat{b}^2}\right],
\label{sss}
\end{eqnarray}
where the coefficients $\beta_{i}$ and $A^{(i)}$ are
\begin{eqnarray}
& &\beta_{1}=\frac{33-2n_{f}}{12}\;,\;\;\;\beta_{2}=\frac{153-19n_{f}}{24}\; ,
\nonumber \\
& &A^{(1)}=\frac{4}{3}\;,
\;\;\; A^{(2)}=\frac{67}{9}-\frac{\pi^{2}}{3}-\frac{10}{27}n_
{f}+\frac{8}{3}\beta_{1}\ln\left(\frac{e^{\gamma_E}}{2}\right)\; ,
\end{eqnarray}
with $\gamma_E$  the Euler constant.

The running coupling constant $\alpha_s$ up to next-to-leading-log
is written as
\begin{equation}
\frac{\alpha_s(\mu)}{\pi}=\frac{1}{\beta_1\ln(\mu^2/\Lambda^2)}-
\frac{\beta_2}{\beta_1^3}\frac{\ln\ln(\mu^2/\Lambda^2)}
{\ln^2(\mu^2/\Lambda^2)}\;.
\end{equation}

The exponent $s(x, b, Q)$ is obtained under the condition that
$xQ/\sqrt 2>1/b$, i.e. the longitudinal momentum should be larger
than the transverse degree. So $s(x, b, Q)$ is defined for
${\hat q}\ge {\hat b}$, and set to zero for ${\hat q}<{\hat b}$.
The previous formulas \cite{BottsSterman} about the exponent $s(x, b, Q)$
picks up the most important first six terms in the first and second
lines of the expression of $s$. Note that the sign of the the fifth
and sixth terms are different from those in \cite{BottsSterman}.

The Sudakov form factor factor $e^{-s(x,b,Q)}$ falls off quickly
in large $b$ region and vanishes as $b>1/\Lambda_{QCD}$. Therefore
it suppresses the long-distance contribution, which is called
Sudakov suppression. In axial-gauge, the Sudakov form factor is
included in each hadronic wave function. So we can define the pion
and B hadronic wave functions with Sudakov corrections as %
\beq
 \Psi_{\pi}(x, b_{\pi}; t)=\exp(-S_{\pi})\Psi^0_{\pi}(x,b_{\pi},t),
   ~~~~~~
 \Psi_{B}(\xi, b_{B}; t)=\exp(-S_{B})\Psi^0_{B}(\xi,b_{B},t),
\eeq %
where $t$ is the factorization scale in the hard scattering
kernel. The $\Psi^0_{\pi}(x,b_{\pi},t)$ and
$\Psi^0_{B}(\xi,b_{B},t)$ are wave functions without Sudakov
corrections.

Combining with the evolution of wave functions and hard scattering
kernel, a complete factor $e^{-S_{\pi, B}}$ for pion and B mesons
can be given as \beq \label{eq:Scom}
 S_{\pi} &=& s(x, b_{\pi}, m_B)+s(\bar x, b_{\pi}, m_B)
   -\frac{1}{\beta_1}\ln\frac{\ln(t/\lqcd)}{\ln(1/(b_{\pi}\lqcd))},\non \\
 S_{B} &=& s(\xi, b_{B}, m_B)
   -\frac{1}{\beta_1}\ln\frac{\ln(t/\lqcd)}{\ln(1/(b_B\lqcd))}.
\label{e37}
\eeq

About the Sudakov form factor  $e^{-s(x,b,m_B)}$ for B and pion meson,
some comments are in order:
\begin{itemize}
 \item The Sudakov factor for B meson is only associated with the
  light quark since there is no collinear divergence associated with
  the heavy $b$ quark. Due to the fact that the momentum of light spectator
  quark is concentrated at the order of $\lqcd$, we may expect the Sudakov
  effects is small because of the suppression of B meson wave function. Our
  numerical analysis shows that its effect is at one percent level.
 \item Notice that the effect of the evolution term in the last of
  eq.(\ref{e37}) gives slight enhancement. If Sudakov suppression
  takes place, the enhancement will be masked ( for example, the
  case of pion ). On the other hand, if Sudakov suppression is not
  effective ( for example, the case of $B$ meson), this enhancement
  will emerge.
 \item In the endpoint region, the exchanged gluon carries small longitudinal
  momentum. The transverse degree cannot be neglected. The mechanism of
  Sudakov suppression begin to take into effect, and the long-distance
  dynamics from large transverse separations are suppressed by Sudakov
  effects. If the momentum fraction of one quark $x$ is small, the Sudakov
  form factor associated with it is 1, no suppression takes place.
  However the Sudakov form factor associated with the other quark
  will provide strong suppression because the momentum fraction of this
  quark $\bar x$ is large now.
\end{itemize}

%%%%%%%%%%%%%%%%%%%%%%%%%%%%%%%%%%%%%%%%%%%%%
%   Threshold resummation
%%%%%%%%%%%%%%%%%%%%%%%%%%%%%%%%%%%%%%%%%%%%%
\subsection{ Threshold resummation }

Now, we discuss the threshold resummation effects. Double
logarithms $\alpha_s\ln^2 x$ can be produced by the higher order
loop corrections. It diverges at the endpoint. If the endpoint
contribution is important, the double logarithms $\alpha_s\ln^2 x$
needs to be resumed to all orders. We review the basic idea of
threshold resummation. Our discussion underlies on the analysis
given in \cite{Liform,Lithreshold}.  The vertex corrections at
$\alpha_s$ order produce the double logarithms
$-\frac{\alpha_s}{4\pi}C_F\ln^2 x$ where $C_F=4/3$ is the color
factor. This collinear divergences can be factorized into a quark
jet function $S_t(x)$. In order to resume the double logarithms to
all orders, it is necessary to introduce the moment (N) space. In
N space, the Sudakov factor has the exponential form up to the
accuracy of leading-log approximation (LL), \beq
S_t^{(LL)}(N)=\exp\left[-\frac{1}{4}\gamma_K^{(LL)}\ln^2 N\right],
\eeq where the anomalous dimension
$\gamma_K^{(LL)}=\alpha_sC_F/\pi$.

The jet funation $S_t(x)$ can be obtained from $S_t(N)$ through the
transfromation
\beq
S_t(x)=\int_{a-i\infty}^{a+i\infty}\frac{dN}{2\pi i}(1-x)^{-N}
  S_t(N)S_t^{(0)}(N)\;,
\label{th1}
\eeq
where $a$ is an arbitrary real constant larger than all the real parts of
poles involved in the integrand. The $S_t^{(0)}(N)$ comes from Mellin
transformation of the initial condition $S^{(0)}_t(x)=1$,
\beq
S_t^{(0)}(N)=\int_0^1 dx (1-x)^{N-1}S^{(0)}_t(x)=\frac{1}{N}\;.
\eeq
The upper index ``0" means that there is no QCD corrections.

The contour integral in Eq.~(\ref{th1}) can be transformed to
\beq \label{St(x)}
S_t^{(LL)}(x)=-\exp\left(\frac{\pi}{4}\alpha_sC_F\right)
 \int_{-\infty}^{\infty}\frac{dt}{\pi}(1-x)^{\exp(t)}
 \sin\left(\frac{1}{2}\alpha_s C_Ft\right)
 \exp\left(-\frac{\alpha_s}{4\pi}C_Ft^2\right)\;.
\label{th2}
\eeq

The above analysis can be directly extended to next-to-leading
logarithms. The jet function $S_t(x)$ satisfies the normalization
condition $\int_0^1 S_t(x)=1$. It vanishes at the end points $x\to
0$ and $x\to 1$. The most important property of jet function
$S_t(x)$ is that it damps faster than any power of $x$.

Since the resumed factor $S_t(x)$ suppresses small $x$ contribution, it may
play crucial role in B decays. The B meson distribution amplitude
$\phi^-_B(x)$ and the twist-3 distribution amplitudes $\phi_P(x)$ do not
vanish at $x=0$ in general. Although the transverse momentum can regulate
the endpoint singularity,  there is still a substantial contribution coming
from endpoint region. The factor $S_t(x)$ can suppress endpoint contribution
and make pQCD more applicable.

In \cite{Yao}, the authors discuss another resummation whose formula is
similar to Sudakov resummation. The obtained Sudakov form factor
suppresses small $x$ contribution more rapidly than the factor $S_t(x)$.
It shows the importance of double-log corrections from another point.

The factor $S_t(x)$ presented in Eq.(\ref{St(x)}) involves one parameter
integration. In order to simplify the numerical calculation, a simple
parametrization for $S_t(x)$ is proposed \cite{Liform}
\beq  \label{eq:sts}
S_t(x)=\frac{2^{1+2c}\Gamma(3/2+c)}{\sqrt{\pi}\Gamma(1+c)} [x(1-x)]^c\;,
\label{stx}
\end{eqnarray}
where the parameter $c$ is determined around $0.3$. The factor $S_t(x)$ with
the above simple parameterization form vanishes at $x=0, 1$. But it does not
damp faster than any power of $x$. Thus the above parametrization is proposed
only for phenomenological application. The rigorous treatment should retain
the integral.

%%%%%%%%%%%%%%%%%%%%%%%%%%%%%%%%%%%%%%%%%%%%%%%%%%%%%%
%   The $B\to\pi $ form factors in large $m_b$ limit
%%%%%%%%%%%%%%%%%%%%%%%%%%%%%%%%%%%%%%%%%%%%%%%%%%%%%%
\section { Calculations}

\subsection { The formulas of $B\to\pi $ form factors in pQCD approach}

We have defined \bpi matrix element in terms of form factors
$F_{+,0}^{B\pi}$ in Eq.~(\ref{eq:bpi1}). The \bpi matrix element
can also be expressed in another form %
\beq \label{eq:bpi2}
 \langle \pi(P_{\pi})|\bar u \gamma_{\mu}b|\bar B(P_B)\rangle=
  f_1(q^2)P_{B\mu}+f_2(q^2)P_{\pi\mu},
\eeq %
From Eq.~(\ref{eq:bpi1}) and Eq.~(\ref{eq:bpi2}), we can
obtain $F_{+,0}^{B\pi}$ from $f_{1,2}$ \beq
  F_+^{B\pi}=\frac{1}{2}(f_1+f_2), ~~~~
  F_0^{B\pi}=\frac{1}{2}(f_1+f_2)+\frac{1}{2}\bar\eta(f_1-f_2)~.
\eeq
with $\bar\eta=\frac{q^2}{m_B^2}$.

\begin{figure}[ht]
 \begin{center}
 \epsfig{file=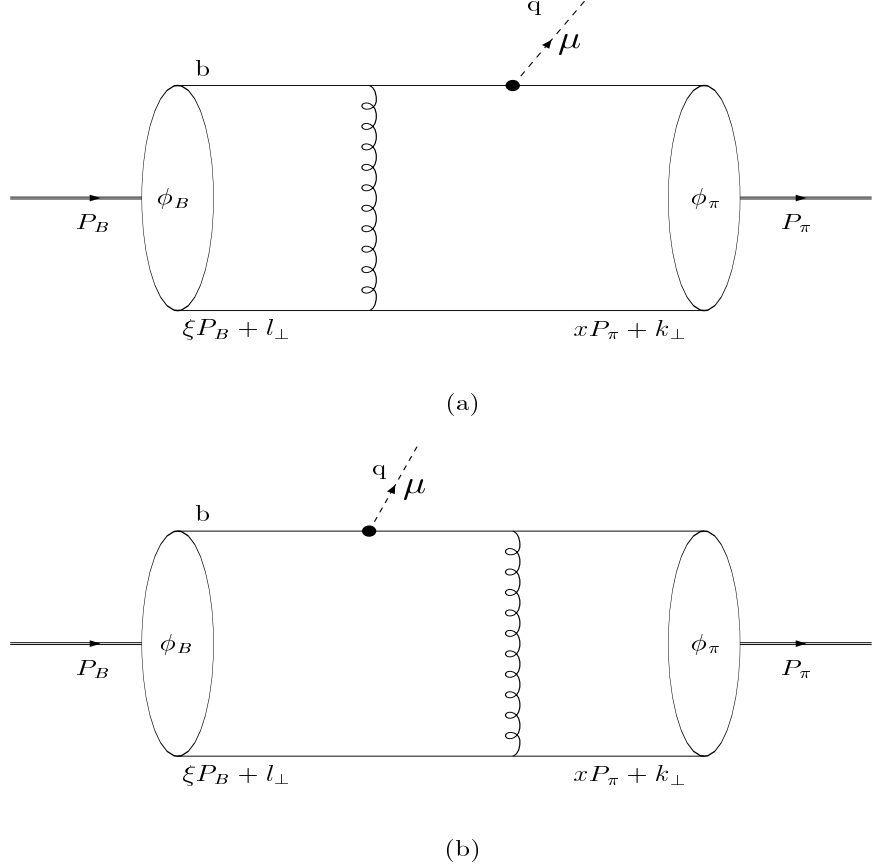, width=14cm,height=13cm}
 \vspace{-1.5cm}
 \end{center}
\caption{ Lowest order hard-scattering kernel for $B\pi$ form
factor}
\label{figbpi}
\end{figure}

In the large recoil region, the \bpi transition is dominated by the single
gluon exchange in the lowest order as depicted in Fig. \ref{figbpi}. The
formulas for the amplitude of Fig. \ref{figbpi}(a) and (b) are
\beq
A&=&\frac{\pi C_F}{N_c} f_{\pi}f_Bm_B^2\int d\xi dx
   \int d^2\pv{l} \int d^2\pv{k}\alpha_s  \frac{1}
   {(x\eta m_B^2+\pv{k}^2)(\xi x\eta m_B^2+(\pv{l}-\pv{k})^2)}
\non \\
 &\times & 2 \Bigg\{
  \Psi_{\pi}\left[(x\eta+1)\Psi_B+(x\eta-1)\bar\Psi_B
   +\frac{\Delta(\xi,l_{\bot})}{m_B}
    \frac{\vec k_{\bot}\cdot(\vec l_{\bot}-\vec k_{\bot})}
     {\xi x\eta m_B^2+(\pv{l}-\pv{k})^2} \right] P_{\pi\mu}
\non \\
 && ~~~+\frac{\mu_\pi}{m_B}\Psi_p \left[(\Psi_B-\bar \Psi_B)P_{B\mu}
  +2(\frac{1}{\eta}\bar\Psi_B-x\Psi_B)P_{\pi\mu} \right]
\non \\
 && ~~~-\frac{\mu_\pi}{m_B}\frac{\Psi'_{\sigma}}{6}\left[
  (\Psi_B-\bar \Psi_B)P_{B\mu}
  -2(\frac{1}{\eta}-x)\Psi_B P_{\pi\mu} \right]
 \\
 && ~~~-2\frac{\mu_\pi}{m_B}\frac{\Psi_{\sigma}}{6} \Bigg [
  2\left( -1+\frac{\pv{k}^2}{x\eta m_B^2+\pv{k}^2}
    -\frac{\pv{k}\cdot(\pv{l}-\pv{k})}
     {\xi x\eta m_B^2+(\pv{l}-\pv{k})^2} \right)\Psi_B
\non \\
 && ~~~+\left( -1-\frac{\pv{k}\cdot(\pv{l}-\pv{k})}{x\eta m_B^2+\pv{k}^2}
  +\frac{2(\pv{l}-\pv{k})^2}{\xi x\eta m_B^2+(\pv{l}-\pv{k})^2} \right)
    \frac{m_B\Delta(\xi,l_\bot)}{\xi x\eta m_B^2+(\pv{l}-\pv{k})^2}
    \Bigg ] P_{\pi\mu}
  \Bigg\},
\non
\eeq
and
\beq
B&=&\frac{\pi C_F}{N_c} f_{\pi}f_Bm_B^2\int d\xi dx
   \int d^2\pv{l} \int d^2\pv{k} \alpha_s \frac{1}
   {(\xi\eta m_B^2+\pv{l}^2)(\xi x\eta m_B^2+(\pv{l}-\pv{k})^2)}
\non \\
 &\times&  2 \Bigg\{
  \Psi_{\pi}\Bigg [ \xi\eta(\Psi_B+\bar\Psi_B)P_{B\mu}-
  \xi(\Psi_B+\bar\Psi_B)P_{\pi\mu} %\non
\\
&& ~~~~~~~~+\frac{\Delta(\xi,l_\bot)}{m_B}
 \left(1-\frac{\pv{l}^2}{\xi\eta m_B^2+\pv{l}^2}
  -\frac{(\pv{l}-\pv{k})\cdot\pv{l}}{\xi x\eta m_B^2+(\pv{l}-\pv{k})^2}
 \right)P_{\pi\mu} \Bigg ]
\non \\
&& +2\frac{\mu_\pi}{m_B}\Psi_p  \Bigg [\left(
 \xi(\bar\Psi_B-\Psi_B)+\frac{\Delta(\xi,l_\bot)}{m_B}
 (1-\frac{\pv{l}^2}{\xi\eta m_B^2+\pv{l}^2}
  -\frac{(\pv{l}-\pv{k})\cdot\pv{l}}{\xi x\eta
  m_B^2+(\pv{l}-\pv{k})^2}) \right)P_{B\mu}
\non \\
&& ~~~~~~~~~~~~~~+\left ( (1-\frac{2}{\eta}\xi)\bar\Psi_B+\Psi_B \right )
 P_{\pi\mu} \Bigg ] \Bigg\}.
\non
\eeq

In hard scattering kernels, transverse momentum $k_\bot$ in the
denominators are retained to regulate the endpoint singularity.
The $k_\bot^2$ in the numerator are power suppressed compared to
$m_B^2$ and they must be dropped in order to ensure the gauge
invariance required by factorization theorem. Transform the
formulas from the momentum space into transverse configuration
$b$-space and include Sudakov factors, we obtain the final
formulas for $F_+^{B\pi}$ and $F_0^{B\pi}$ in pQCD approach \beq
\label{eq:f+} F_+^{B\pi} &=& \frac{\pi C_F}{N_c}
f_{\pi}f_Bm_B^2\int d\xi dx
     \int b_Bdb_B~ b_\pi db_\pi~ \alpha_s(t) \\
&\times& \Bigg \{
 \Bigg [ \Psi_\pi(x, b_\pi)\left ( (x\eta+1)\Psi_B(\xi, b_B)+
  (x\eta-1)\bar\Psi_B(\xi, b_B) \right )
\non \\
&& +\frac{\mu_\pi}{m_B}\Psi_p(x, b_\pi) \left(
  (1-2x)\Psi_B(\xi,b_B)+(\frac{2}{\eta}-1)\bar\Psi_B(\xi,b_B)\right)
\non \\
&& -\frac{\mu_\pi}{m_B}\frac{\Psi'_\sigma(x,b_\pi)}{6} \left(
 (1+2x-\frac{2}{\eta})\Psi_B(\xi,b_B)-\bar\Psi_B(\xi,b_B) \right)
+4\frac{\mu_\pi}{m_B}\frac{\Psi_\sigma(x,b_\pi)}{6}\Psi_B(\xi,b_B)
 \Bigg ]h_1
\non \\
&& -2\frac{\mu_\pi}{m_B}\frac{\Psi_\sigma(x,b_\pi)}{6}
 (-m_B\Delta(\xi,b_B))\frac{b_B}{2\sqrt{\xi x\eta}m_B}h_3
\non \\
&& +\Bigg [ \Psi_\pi(x, b_\pi)\left ( -\xi\bar\eta
 (\Psi_B(\xi,b_B)+\bar\Psi_B(\xi,b_B))+\frac{\Delta(\xi,b_B)}{m_B}
 \right )
\non \\
&& +2\frac{\mu_\pi}{m_B}\Psi_p(x,b_\pi)\left ( (1-\xi)\Psi_B(\xi,b_B)
  +(1+\xi-\frac{2\xi}{\eta})\bar\Psi_B(\xi,b_B)
  +\frac{\Delta(\xi,b_B)}{m_B}\right ) \Bigg ]
 h_2%(\xi,x,b_B,b_B)
\Bigg \},
\non
\eeq
and
\beq \label{eq:f0}
F_0^{B\pi} &=& \frac{\pi C_F}{N_c} f_{\pi}f_Bm_B^2\int d\xi dx
     \int b_Bdb_B~ b_\pi db_\pi~ \alpha_s(t) \\
&\times& \Bigg \{
 \Bigg [ \Psi_\pi(x, b_\pi)\eta\left ( (x\eta+1)\Psi_B(\xi, b_B)+
  (x\eta-1)\bar\Psi_B(\xi, b_B) \right )
\non \\
&& +\frac{\mu_\pi}{m_B}\Psi_p(x, b_\pi) \left(
  (2-\eta-2x\eta)\Psi_B(\xi,b_B)+\eta\bar\Psi_B(\xi,b_B)\right)
\non \\
&& -\frac{\mu_\pi}{m_B}\frac{\Psi'_\sigma(x,b_\pi)}{6} \left(
 \eta(2x-1)\Psi_B(\xi,b_B)-(2-\eta)\bar\Psi_B(\xi,b_B) \right)
+4\frac{\mu_\pi}{m_B}\eta\frac{\Psi_\sigma(x,b_\pi)}{6}\Psi_B(\xi,b_B)
 \Bigg ]h_1
\non \\
&& -2\frac{\mu_\pi}{m_B}\frac{\Psi_\sigma(x,b_\pi)}{6}
 (-m_B\Delta(\xi,b_B))\frac{b_B}{2\sqrt{\xi x\eta}m_B}h_3
\non \\
&& +\Bigg [ \Psi_\pi(x, b_\pi)\eta\left ( \xi\bar\eta
 (\Psi_B(\xi,b_B)+\bar\Psi_B(\xi,b_B))+\frac{\Delta(\xi,b_B)}{m_B}
 \right )
\non \\
&& +2\frac{\mu_\pi}{m_B}\Psi_p(x,b_\pi)\left (
  (\eta-2\xi+\xi\eta)\Psi_B(\xi,b_B)
  +\eta(1-\xi)\bar\Psi_B(\xi,b_B)
  +(2-\eta)\frac{\Delta(\xi,b_B)}{m_B}\right ) \Bigg ] h_2
\Bigg \},
\non
\eeq
where
\beq  \label{eq:h}
% h_1
& h_1=K_0(\sqrt{\xi x\eta}~m_B b_B)
    \Bigg [ \theta(b_B-b_\pi)I_0(\sqrt{x\eta}~m_Bb_\pi)
      K_0(\sqrt{x\eta}~m_B b_B) \non \\
&  ~~~~~~~~~~~~~~~~~~~~~~~~~~~~~~~~~~~
 +\theta(b_\pi-b_B)I_0(\sqrt{x\eta}~m_Bb_B)
    K_0(\sqrt{x\eta}~m_B b_\pi) \Bigg ], \\
% h_2
& h_2=K_0(\sqrt{\xi x\eta}~m_B b_\pi)
    \Bigg [ \theta(b_B-b_\pi)I_0(\sqrt{\xi\eta}~m_Bb_\pi)
      K_0(\sqrt{\xi\eta}~m_B b_B) \non \\
&  ~~~~~~~~~~~~~~~~~~~~~~~~~~~~~~~~~~~
 +\theta(b_\pi-b_B)I_0(\sqrt{\xi\eta}~m_Bb_B)
    K_0(\sqrt{\xi\eta}~m_B b_\pi) \Bigg ], \\
% h_3
& h_3=K_{-1}(\sqrt{\xi x\eta}~m_B b_B)
    \Bigg [ \theta(b_B-b_\pi)I_0(\sqrt{x\eta}~m_Bb_\pi)
      K_0(\sqrt{x\eta}~m_B b_B) \non \\
&  ~~~~~~~~~~~~~~~~~~~~~~~~~~~~~~~~~~~
 +\theta(b_\pi-b_B)I_0(\sqrt{x\eta}~m_Bb_B)
    K_0(\sqrt{x\eta}~m_B b_\pi) \Bigg ].
\eeq
The function $K_i$ and $I_i$ are modified Bessel functions with $i$ their
orders.

The physical quantities should not depend on the choice of the
scale parameter $t\equiv\mu$ if the calculation can be performed
up to infinite orders. Therefore,  the scale parameter can be
chosen as any value in principle. However, in practice the
calculation can only be made perturbatively. To make the
perturbative expansion meaningful, the scale parameter should be
chosen in such a way that can make the higher order corrections
small. The natural choice is $t = \sqrt{\xi x \eta}~m_B$ in the
standard approach. If $\xi, x\to 0$, $\alpha_s(t)$ will be
divergent at $t\leq\lqcd$. When including the transverse degree of
freedom, $\alpha_s (\mu) \ln \sqrt{\xi x \eta}~m_B/\mu$, $\alpha_s
(\mu) \ln b_B\mu$ and $\alpha_s (\mu) \ln b_\pi\mu$ will appear in
higher order corrections. Therefore we take $t={\rm max}(\sqrt{\xi
x \eta}m_B, 1/b_B, 1/b_\pi)$. The scale $t\equiv\mu$ must be
larger than $\lqcd$ thus avoids the divergence of coupling
constant.

The wave functions include Sudakov corrections coming from Sudakov and
threshold resummations
\beq
\Psi_\pi(x,b_\pi)&=&S_t(x)\exp(-S_\pi)\phi_{\pi}(x)\Sigma_\pi(b_\pi), \\
\Psi_B(\xi,b_B)&=&S_t(\xi)\exp(-S_B)\phi_B(\xi)\Sigma_B(b_B). \non
\eeq
The similar expressions are given for $\Psi_p, \Psi_\sigma, \Psi'_\sigma,
\bar\Psi_B$. The jet function $S_t$ comes from threshold resummation. The
simplified parametrization form in Eq.(\ref{eq:sts}) is taken to estimate
threshold resummation effects in this paper. The complete Sudakov factors
$S_{\pi,B}$ are given in Eq.(\ref{eq:Scom}).  $\Sigma_{\pi,B}$ are
intrinsic transverse momentum dependence of pion and B meson wave functions.

We compare our formulas with the results in
\cite{Sachrajda,Liform}. In \cite{Sachrajda}, only the leading
twist of pion is discussed. Set the twist-3 terms to zero, the two
formulas of ours and \cite{Sachrajda} are the same except the
definition of $h_1$. The difference comes from the Fourier
transform of hard part. In \cite{Liform}, the single B meson wave
function $\Psi_B$ is assumed and the terms of $\bar\Psi_B$ and
$\Delta$ are neglected. The twist-3 power correction is included.
The momentum projector in \cite{Liform} for pion meson is slightly
different from our projector in Eq.(\ref{eq:pionprojector}).
Except for these differences, the formulas in \cite{Liform} are
consistent with ours.

\subsection {The assumption of single distribution amplitude}

Before we perform analysis of \bpi form factors in pQCD approach,
we discuss an assumption of using single B meson distribution
amplitude at first. To our knowledge, the use of B meson
distribution amplitude in B decays firstly appeared in
\cite{Brodsky}. The authors suggest the simplest momentum
projection for B meson which contains two terms %
\beq
 M_B=-\frac{if_B}{4}[\dirac p_B+m_Bg(\xi)]\gamma_5\phi_B(\xi).
\eeq%
The function $g(\xi)$ is assumed to be at the order of 1. Setting
$g(\xi)=1$, the B meson momentum projection reduces to \beq
 M_B=-\frac{if_B}{4}[\dirac p_B+m_B]\gamma_5\phi_B(\xi).
\eeq This is the widely used formula in previous pQCD analysis
which contains single B meson distribution amplitude. Here, we
discuss the most familiar model used in the previous pQCD analysis
\beq \label{eq:phib}
 \phi_B(\xi)=N_B\xi^2(1-\xi)^2\exp \left ( -\frac{\xi^2 m_B^2}
  {2\omega_B^2} \right ),
\eeq where $N_B$ is the renormalization constant makes
$\int_0^1\phi_B(\xi)d\xi=1$. In this model $\phi_B(\xi)$ has a
peak at $\xi=\bar\Lambda/m_B$ where $\bar\Lambda= m_B-m_b$. This
model of B meson distribution amplitude has been used to fit
different channels of B decays.

B meson distribution amplitudes are important ingredients in pQCD
approach. We should be careful about the choice of B meson
distribution amplitude. In HQET, the definition of B meson
distribution amplitude contains two terms $\phi_B^+,~ \phi_B^-$.
Compare Eq.(\ref{eq:phib}) and Eq.(\ref{eq28}), the difference
lies in $1/m_B$ effect and can be neglected. The choice of single
distribution amplitude amounts to taking $\phi_B=\phi_{B1}$ and
neglects $\phi_{B2}$. So, the validity of the assumption of single
distribution amplitude depends on whether the contribution of
$\phi_{B2}$ is small or not. To test this assumption, we make
approximation that $\phi_B\approx \phi_{B1}=\phi_B^+$ where
$\phi_B^+$ is given in Eq.(\ref{eq28}). This approximation is
reasonable because the calculated form factor $F^{B\pi}(0)$ is:
0.239 using $\phi_B$ in Eq.(\ref{eq:phib}) and 0.227 using
$\phi_B^+$ in Eq.(\ref{eq28}).

Table 1 shows the numerical result for the contributions from
$\phi_{B1}, \phi_{B2}$ and $\Delta_B$.
\begin{table}[hbt] \label{taphibbar}
\begin{center}
\parbox{14cm}{\caption{The \bpi form factors $F^{B\pi}=F_{+,0}^{B\pi}(0)$ with
$\phi_{B1}$, $\phi_{B2}$ and $\Delta_B$. The column
``$\phi_{B1}$'' represents $\phi_{B1}$ contribution only. The
meanings of columns ``$\phi_{B2}$'' and ``$\Delta_B$'' are
similar.  The column ``sum'' represents the summation of all these
contributions. }}
\begin{tabular}{|c|c|c|c|c|} \hline \hline
    & $\phi_{B1}$ & $\phi_{B2}$ &  $\Delta_B$ & sum \\ \hline
twist-2 & 0.042   & 0.016       &  0.014      & 0.072 \\ \hline
twist-3 & 0.185   & 0           &  0.048      & 0.233 \\ \hline
total   & 0.227   & 0.016       &  0.062      & 0.305\\
\hline\hline
\end{tabular}
\end{center}
\end{table}
If considering only $\phi_{B1}$ contribution, $F^{B\pi}(0)$ is
0.227. This result is consistent with the one using QCD sum rule
within the theoretical errors. The $\phi_{B1}$ contribution is
dominant. For $\phi_{B2}$ term, its contribution vanish at twist-3
but cannot be neglected at twist-2. The power suppressed term of
$\Delta_B$ is about 20\% in the total numerical result. From a
general point,  there should have two leading B meson distribution
amplitudes (or generally wave functions) in pQCD framework. But
the single B meson distribution amplitude $\phi_{B1}$ is the
dominant contribution. It should be noted that $\phi_{B2}$
contribution vanishes in the hard spectator scattering and weak
annihilation diagrams in $B\to \pi\pi$ decays.

\subsection { The \bpi form factors for different models of
distribution amplitudes }

The distribution amplitudes for pion depends on the renormalization scale.
This dependence is controlled by the evolution equation. For the scale related
to our discussion, evolution effect should be important but precise estimate
of this effect depends on the unknown input parameters. In this paper, what we
concern most is B meson wave functions and the reliability of pQCD framework.
We will not consider the evolution effects of pion distribution amplitudes.
Thus, the pion distribution amplitudes for both the twist-2 and twist-3 are
taken as their asymptotic form for simplicity, i.e.,
\beq
 \phi_\pi=6x\bar x,~~~~~~~~~~~ \phi_p=1,~~~~~~~~~~~
 \phi_\sigma=6x\bar x.
\eeq

For B meson distribution amplitudes (wave functions), three models
exist in literatures:
\begin{itemize}
 \item   Model I
  \beq
    \phi^-_B(\xi)=\sqrt\frac{2}{\pi}\frac{m_B}{\omega_B}
     \exp(-\frac{\xi^2 m_B^2}{2 \omega_B^2}), \qquad
     \phi^+_B(\xi)=\sqrt{\frac{2}{\pi}}
     \frac{\xi^2 m_B^3}{\omega_B^3}
     \exp(-\frac{\xi^2 m_B^2}{2 \omega_B^2}),
  \eeq
 \item Model II
  \beq
   \phi^-_B(\xi)=\frac{m_B}{\omega_0}
    \exp(-\frac{\xi m_B}{\omega_0}), \qquad
   \phi^+_B(\xi)=\frac{\xi m_B^2}{\omega_0^2}
    \exp(-\frac{\xi m_B}{\omega_0}),
  \eeq
where $\omega_0=\frac{2}{3}\bar\Lambda$ and $\bar\Lambda=m_B-m_b$
in this model.
 \item Model III
  \beq
    \Psi^-_B(\xi,k_\perp)=\frac{2\bar\xi-\xi}{2\pi\bar\xi^2}\theta(2\bar\xi-\xi)
        \delta(k_\perp^2-m_B^2\xi(2\bar\xi-\xi)),
   \nonumber \\
    \Psi^+_B(\xi,k_\perp)=\frac{\xi}{2\pi\bar\xi^2}\theta(2\bar\xi-\xi)
     \delta(k_\perp^2-m_B^2\xi(2\bar\xi-\xi))~~~,
  \eeq
with $\bar\xi=\frac{\bar\Lambda}{m_B}$.
\end{itemize}

Model I is proposed in \cite{Sachrajda}. It uses the equations of
motion for light spectator quark in HQET. The $\omega_B$ is at the
order of $\lqcd$. The possible range is $0.2\GeV-0.5\GeV$. Model
II is based on a QCD Sum rule inspired analysis
\cite{GrozinNeubert}. The above two models are both Gaussian type
and there is a peak near $\bar\Lambda/m_B$. Model III uses the
equations of motion for both light spectator quark and heavy b
quark in HQET \cite{Qiao}. The distribution amplitudes have a
cutoff at $\xi=2\bar\xi$ because it takes the approximation
$m_B=\infty$ \footnote{We thank J. Kodaira for the discussions
about this topic.}. The distributions are linear functions of
$\xi$. For model I, the transverse momentum dependent function
$\Sigma_B(b_B)$ is given in Eq.(\ref{eq:ktb}). For model II, the
transverse momentum dependence part is unknown, we take the same
transverse momentum dependence function $\Sigma_B(b_B)$ as model
I.

The other input parameters are as follows: decay constants for
pion and B meson $f_\pi=0.13\GeV$, $f_B=0.19\GeV$;
$\lqcd=0.25\GeV$; $\bar\Lambda=m_B-m_b=0.5\GeV$; the oscillator
parameter in the transverse momentum distribution of pion wave
function $\beta^2=4\GeV^{-2}$ \cite{Jakob}; parameter in B meson
wave function $\omega_B=0.35\GeV$; pion twist-3 coefficient
$\mu_\pi=2.2\GeV$ \cite{BBNS2}.

We present the result of form factors $F_{+,0}^{B\pi}$ at large recoil
$q^2=0$. Using B and pion wave functions and the input parameters
presented above, predictions for the $F_{+,0}^{B\pi}(0)$ with the three
models for B meson distribution amplitudes are listed in Table 2.
%\ref{ta:3model}

\begin{table}[hbt] \label{ta:3model}
\begin{center}
\parbox{14cm}{\caption{ The numerical results of \bpi form factors
 $F_{+,0}^{B\pi}(0)$ with three models for B meson distribution amplitudes. }}
\begin{tabular}{|c|c|c|c|c|}  \hline \hline
   Model  & $\phi_B$ & $\bar\phi_B$ &  $\Delta_B$ & total \\ \hline
    I     &  0.564   &  -0.321      &  0.062     & 0.305 \\ \hline
    II    &  0.641   &  -0.338      &  0.060     & 0.363 \\ \hline
    III   &  0.540   &  -0.335      &  0.055     & 0.260\\ \hline
    \hline
\end{tabular}
\end{center}
\end{table}

The predicted \bpi form factors $F_{+,0}^{B\pi}(0)$ in model I, II
and model III are around $0.3$ which is favored by experiment and
consistent with the prediction by other methods, such as QCD sum
rule \cite{Braun}, BSW model \cite{BSW}. The advantage of model
III is that it only uses the QCD equations of motion. In reality,
it is not a model. There is only one universal phenomenological
parameter $\bar \Lambda$ in wave functions.

At the end of this subsection, we would like to discuss the power
corrections. Table 3 shows that the chirally enhanced twist-3
contribution is numerically larger than leading twist
contribution. The large twist-3 contribution is not consistent
with the assumption of twist expansion. The study of power
corrections should be studied in a more careful and more
systematic way which is beyond the scope of this paper.

\begin{table}[hbt] \label{ta:tt}
\begin{center}
\parbox{14cm}{\caption{ The twist-2 and twist-3 contributions of pion in \bpi
form factor $F^{B\pi}=F_{+,0}^{B\pi}(0)$ with different $\omega_B$. }}
\begin{tabular}{|c|c|c|c|c|c|c|}  \hline \hline
 $\omega_B$(\GeV) & 0.25 & 0.3 & 0.35 & 0.4 & 0.45 & 0.5  \\\hline
 twist-2  & 0.086 & 0.078 & 0.072 & 0.066 & 0.061 & 0.056 \\
 twist-3  & 0.345 & 0.280 & 0.233 & 0.198 & 0.171 & 0.150 \\
 total    & 0.431 & 0.358 & 0.305 & 0.264 & 0.232 & 0.206 \\ \hline\hline
\end{tabular}
\end{center}
\end{table}

\subsection {The reliability of pQCD approach in \bpi form
factors}

Now, we examine  the reliability of pQCD analysis in \bpi form
factors. We choose model I of the B meson distribution amplitudes
for illustration. The conclusions of model II and III are similar.

\begin{figure}[ht]
 \begin{center}
 \epsfig{file=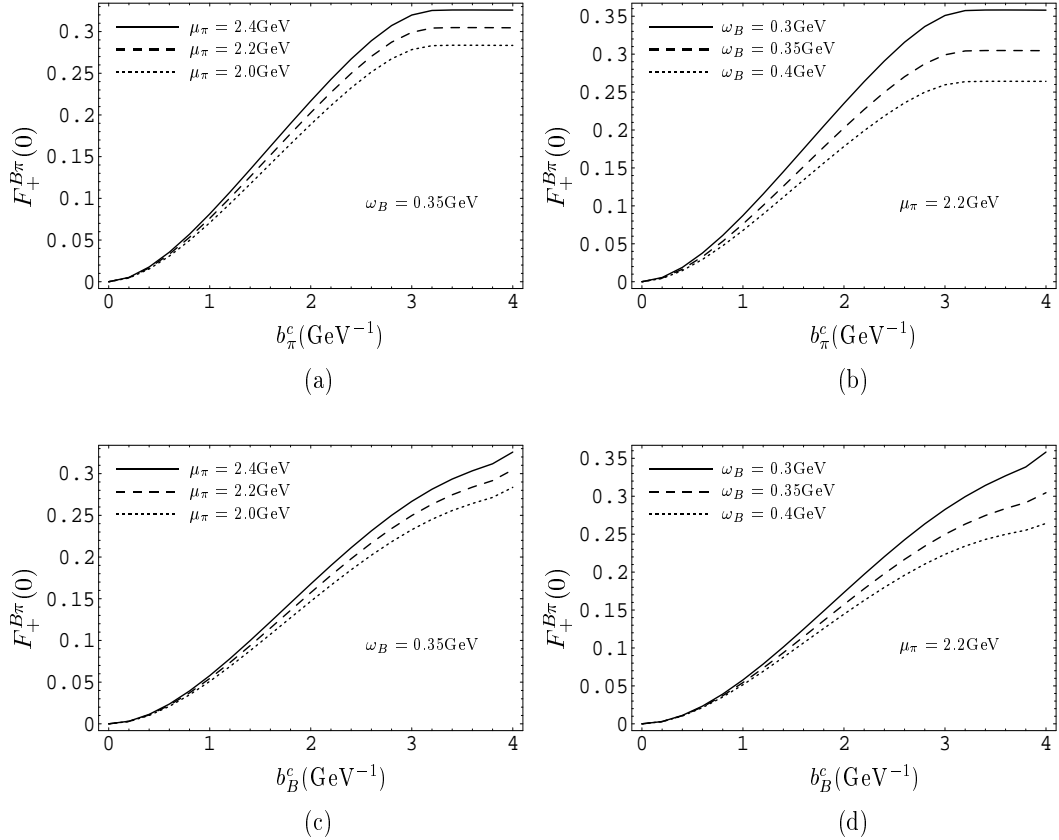, width=16cm, height=13cm}
 \end{center}
\vspace{-1cm}
\caption{$b_\pi$ and $b_B$ dependence of $F^{B\pi}$.} \label{figb2}
\end{figure}

The basic idea of pQCD approach is to use Sudakov effects to
suppress the long-distance contribution with large transverse
separations. A reliable pQCD analysis of \bpi form factors should
satisfy that most of the result comes from the region where impact
parameters $b_\pi, b_B$ are both small. In order to study the
impact parameter $b$ dependence of \bpi form facors, we introduce
a cut-off $b^c$ in impact parameters $b_\pi$ and $b_B$ in the
integrals of Eq. (\ref{eq:f+}) and (\ref{eq:f0}) by
$\int^{b^c_B}_0db_B\int^{b^c_\pi}_0db_\pi$. In Fig. \ref{figb2},
we show the dependence of \bpi form factors
$F^{B\pi}=F^{B\pi}_{+,0}(0)$ on $b^c_\pi$ and $b^c_B$. Fig.
\ref{figb2}(a) plots the $b^c_\pi$ dependence with
$\omega_B=0.35\GeV$ and $\mu_\pi$ varied. Fig. \ref{figb2}(b)
plots the $b^c_\pi$ dependence with $\mu_\pi=2.2\GeV$ and
$\omega_B$ varied. One can see that varying these two important
input parameters $\mu_\pi$ and $\omega_B$ does not change the
behavior of the $F^{B\pi}$ dependence on $b^c$. Similarly, the
$b_B^c$ dependence of $F^{B\pi}$ is depicted in Fig.
\ref{figb2}(c) and (d). From Fig. \ref{figb2}, 50\% of $F^{B\pi}$
comes from the $b^c_\pi<1.5\GeV^{-1}$ for impact parameter $b_\pi$
and $b^c_B<2.0\GeV^{-1}$ for impact parameter $b_B$. The
contributions from regions with large impact parameters $b_\pi,
b_B>2\GeV^{-1}$ are substantial: 31\% for impact parameter $b_\pi$
and 50\% for $b_B$. The calculation of pQCD approach may still
include large long-distance contributions. Therefore, the reliability of
leading order calculations in $\alpha_s$ expansion
should be checked carefully. The more direct criteria is to check the
distribution of the coupling constant $\alpha_s(t)$.

\begin{figure}[ht]
 \begin{center}
 \epsfig{file=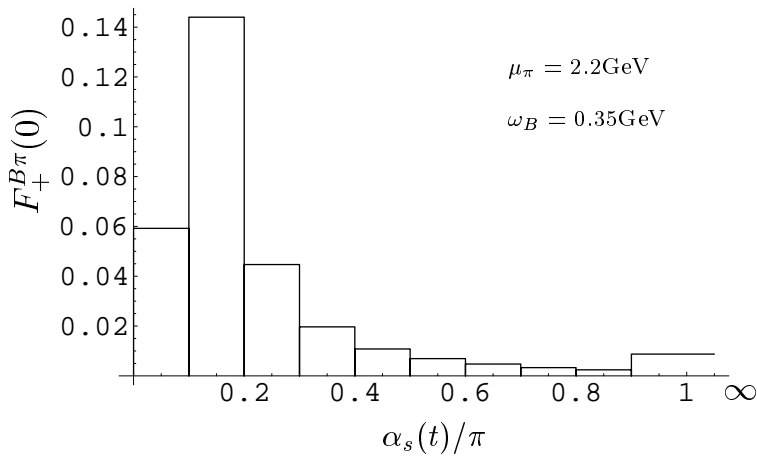, width=12cm}
 %, height=15cm}
   \vspace{-1.5cm}
 \end{center}
\vspace{0.7cm}
\caption{The \bpi form factors $F^{B\pi}$ vs the coupling constant
 $\alpha_s(t)$} \label{fig:as}
\end{figure}

The standard to judge the reliability of
perturbative analysis is that most of the contribution comes from
the region where the coupling constant $\alpha_s(t)$ is small.
Fig. \ref{fig:as} plots the form factors $F^{B\pi}$ coming from
the region where $a_1\leq \alpha_s(t)/\pi\leq a_2$. The last bar
is the contribution of $\alpha_s(t)/\pi>0.9$. In our calculation,
70\% of the result comes from the region where
$\alpha_s(t)/\pi<0.2$ and the
contribution for $t\geq 1\GeV$ is 38\%. If we consider the energy
$1\GeV$ is the point that the perturbation theory is broken, the
non-perturbative contribution is about 60\%. If we consider a
weaker criterion that $\alpha_s(t)/\pi=0.2$ is the broken point,
one can see the non-perturbative contribution is 30\%. No matter
which criterion is chosen, the non-perturbative contribution is
comparable to the perturbative part. So, the prediction of \bpi
form factor in pQCD approach can not be precise. The
non-perturbative contribution constitutes the intrinsic systematic
error in pQCD approach.

\begin{figure}[ht]
 \begin{center}
 \epsfig{file=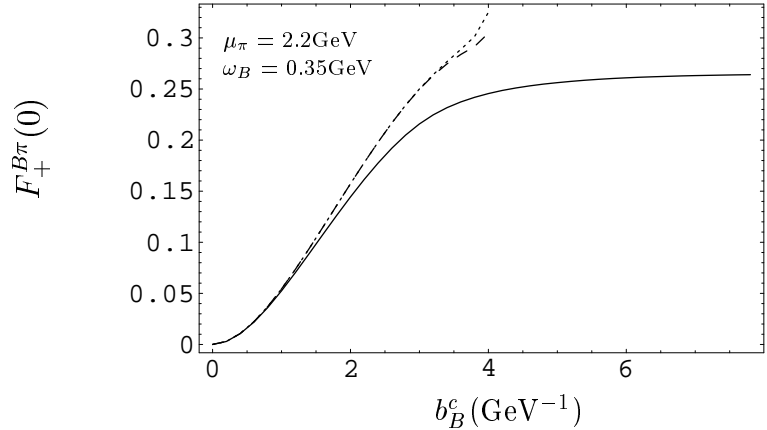, width=12cm}
 %, height=15cm}
   \vspace{-1.5cm}
 \end{center}
\vspace{0.7cm}
\caption{The property of $F^{B\pi}$ vs. $b^c_B$ in B meson. The dotted curve
   is the result with Sudakov and evolution effects in $B$ meson, the
   dashed one is for the case without Sudakov effect, i.e., without
   the contribution of $s(\xi,b_B,m_B)$ in $S_B$, and the solid curve
   is the result without both Sudakov and evolution effects
   in $B$ meson, i.e., without the total contribution of $S_B$.}
 \label{fig:ns}
\label{figsuda}
\end{figure}

One may also wonder if the integration over $b_B$ may not converge
because the form factor $F^{B\pi}_+(0)$, as shown in
Fig.\ref{figb2}(c) and (d), does not saturate at any value of
$b_B^c$. Therefore, it is necessary to study the property of the
dependence of $F^{B\pi}_+(0)$ on $b_B$ in the region
$b_B>1/\lqcd$. However, since the complete factor $e^{-S_B}$ in B
meson wave function is divergent at $\lqcd$, we drop this term and
perform the calculations again. We find that the form factors
saturate at $b_B\sim 1/\lqcd$ as shown in Fig. \ref{fig:ns}. That
is to say, even without any Sudakov suppression, the region of
large $b_B$ does not contribute.

\subsection {Comparison with other works}

Whether \bpi form factors are hard dominant or soft dominant is a
highly controversial topic in B physics. There are many
theoretical investigations using various approaches. In this
paper, we have studied the perturbative method. The comparison of
our result with other recent works in pQCD approach is helpful to
interpret the perturbative method.

In \cite{Liform}, the authors calculate the \bpi form factors in
pQCD approach which includes pion twist-3 power corrections. The
intrinsic transverse momentum dependence of B meson wave function
is considered and of pion is neglected. Neglecting the intrinsic
transverse momentum effect in pion is unjustified. The physical
reason is that Sudakov suppression is not so strong for realistic
B decays and the large $b$ contribution is very sensitive to the
non-perturbative dynamics in hadrons. We have discussed that the
model of B meson wave function chosen in \cite{Liform} is nearly
same as the $\phi_B^+$ in model I. So, we find an explanation of
this model in HQET. The reliability of \bpi form factor in pQCD is
given in \cite{pQCD} by Y. Keum, H. Li and A.I. Sanda. They
observed that 97\% of the contribution comes from the region with
$\alpha_s(t)/\pi<0.3$. This criterion of $\alpha_s(t)/\pi<0.3$ is
not strong enough to guarantee the reliability of leading order
result in $\alpha_s$ expansion series.

The authors in \cite{Sachrajda} address some critical questions
about Sudakov effects in \bpi form factors.
We will make our comments on their questions below.

The first question concerns theoretical issue on Sudakov form
factor. For pion meson, it is pointed out that Sudakov form factor
with next-to-leading-log (NLL) approximation is gauge dependent.
This dependence can be cancelled by a soft function $U(b)$ which
resummed the single soft logarithms \cite{li1997}. A complete
analysis of \bpi form factors should include the
next-to-leading-order (NLO) calculation in order to ensure gauge
invariance. In this paper, we take the NLL Sudakov form factor and
hard kernel to leading order, so the gauge dependence exists in
our analysis. Whether this gauge dependence will lead to large
numerical difference is not clear at present. The NLO calculation
in pQCD approach must include the transverse degrees of freedom,
therefore the calculation is difficult in technique.

The second question is about numerical dependence on the
uncertainties in meson's distribution amplitude. This is an
important and inevitable problem in perturbative method. We find
that the three models of B meson wave functions which satisfy the
QCD equations of motion give the consistent numerical results.
More importantly, the B meson wave functions derived in
\cite{Qiao} is model-independent. So, the theoretical errors
caused by the B meson wave functions can be reduced largely.

The third question is about the reliability of pQCD approach in
\bpi form factors. The conclusion given in \cite{Sachrajda} is
that Sudakov suppression is so weak that pQCD approach cannot be
applied in B physics. After including the intrinsic transverse
momentum effects and threshold resummation, the perturbation
behavior of the numerical result in \cite{Sachrajda} is similar to
ours. Whether the perturbative calculation is reliable or not
should be checked by higher order calculations in $\alpha_s$
expansion series. Our present analysis shows that one can
not guarantee higher order terms will be small. More works are
needed before a completely reliable method is set up to control
the higher order contributions.

\section { Conclusions and discussions }

In this paper, we have presented a systematic analysis of \bpi
form factors in pQCD approach. The intrinsic transverse momentum
effects and threshold resummation are important ingredients in the
pQCD framework in B decays. One important finding in this paper is
that all the three models of B meson wave functions predict
$F^{B\pi}$ about 0.3 in pQCD approach. This result should not be
accidental. We find that the numerical results are not so
sensitive to the choice of B meson wave functions in HQET. The B
meson wave functions which satisfy the QCD equations of motion can
reduce the theoretical errors largely. The consistent numerical
results give us confidence that our knowledge about the B meson
wave functions is not so poor as expected before. HQET provides a
reasonable framework to interpret the intrinsic dynamics in B
meson.

Although the prediction of \bpi form factors in pQCD approach
seems favorable in experiment, we cannot avoid many serious
conceptual problems of pQCD approach. The calculation in our paper
is at the leading order (LO). There is no strong evidence that NLO
result is small especially in our case where most of contribution
comes from the momentum region of $t<1\GeV$. A critical problem in
NLO calculation is how to deal with the dependence on the choice
of renormalization scale $\mu$. In this paper, the scale $\mu$ is
taken as the largest momentum carried by the exchanged gluon.
Whether this choice is best or not depends on the NLO calculation.
However, as we have discussed, the NLO calculation in \bpi form
factors is technically difficult. Another problem comes from power
corrections. We only discuss the chirally enhanced twist-3
correction. There are many other power corrections, such as
twist-4, higher Fock states etc. Although one can argue that they
are power suppressed in the heavy quark limit, we don't know their
numerical effects in the realistic case. In principle, all the
above issues should be treated in a more systematic way than the
present form. The seeming correspondence of the prediction of pQCD
approach in \bpi form factors with other theoretical expectations
is based on many assumptions and one must be careful to obtain a
conclusion. To some extent, we think it is difficult to estimate
the theoretical errors in the present pQCD approach in B decays.

Although there are many crucial problems in pQCD approach, completely
rejecting it should be with caution. There
are at least 30\% hard contribution coming from the moemtnum
$t>1\GeV$ in \bpi form factors. This contribution cannot be
calculated reliably in any other present theoretical approaches.
We cannot throw one method before we get the final answers.

There is no universal, satisfactory criteria of QCD because we
don't know how to quantitatively calculate the non-perturbative
dynamics especially in exclusive, non-leptonic B-meson decays at
present. The complicated strong interactions in exclusive B decays
where both perturbaive and non-perturbative dynamics contribute is
a challenging task for the present non-perturbative methods, such
as QCD sum rules, lattice QCD etc.

In conclusion, \bpi form factors provide an interesting places to
study the perturbative and non-perturbative strong interactions. The
success of B meson wave functions in HQET  is helpful to
understand the exclusive B decays. The perturbative method
provides a simple phenomenological method to estimate the
perturbative aspects of strong dynamics in B decays.

\section*{Acknowledgment}

We would like to thank H. Li, S. Brodsky, A. Ali and V. Braun for
many valuable discussions and their comments. Z. Wei also thanks
M. Beneke and C.T. Sachrajda for the discussions in Regensburg,
20-22 July 2001. M. Yang thanks the partial support of the
Research Fund for Returned Overseas Chinese Scholars. This work is
supported in part by National Natural Science Foundation of China.

\section*{Appendix}

In this appendix we will derive the projector for $B$ meson in the
momentum space with including transverse momentum. We begin with
the generalized Lorentz decomposition of
the light-cone matrix element \cite{GrozinNeubert, Sachrajda}
\beq
M(z)_{\alpha\beta}&\equiv& \langle 0|\bar{q}_\beta(z)
b_\alpha(0)|\bar{B}(p_{B})\rangle \nonumber\\
  &=&-\frac{if_B}{4} \left\{ \frac{\dirac{p}_B+m_B}{2}
    \left[ 2\tilde\Psi^+_B(z^2, t)
    +\frac{\tilde\Psi^-_B(z^2, t)-\tilde\Psi^+_B(z^2, t)}{t}\dirac{z}
    \right] \gamma_5 \right\} _{\alpha\beta},
\label{m1}
\eeq
with $v=\frac{p_B}{m_B}$ and $t=v\cdot z$. In light-cone
coordinate
$z=(\frac{z_+}{\sqrt{2}},\frac{z_-}{\sqrt{2}},z_{\perp})$, with
$z_{\pm}=z_0\pm z_3$. To obtain the projector in the momentum
space we consider the amplitude of one process,
\beq
 T=\int \frac{d^4 z}{(2\pi)^4} M(z) A(z),
\eeq
where $A(z)$ is the hard scattering kernel of the relevant
process. Then
\beq
 T&=&\int \frac{d^4 z}{(2\pi)^4} M(z)\int d^4 l e^{-i l\cdot z} A(l)\nonumber\\
  &=&\frac{1}{4}\int \frac{dz_+dz_-d^2z_{\perp}}{(2\pi)^4} M(z)
  \int dl_+dl_-d^2l_{\perp}
  e^{-i(\frac{l_+z_-+l_-z_+}{2}-\pv{l}\cdot\pv{z})}
  A(l_+,l_-,\pv{l}).
  \label{t1}
\eeq
For the case that $A(l)$ is independent of $l_-$, we have
$A(l)=A(l_+,\pv{l})$. Then the integration over $l_-$ in the
above equation can be accomplished to get a delta function
$\delta (z_+/2)$. Therefore we get
\beq
T=\frac{1}{2}\int \frac{dz_-d^2z_{\perp}}{(2\pi)^3}M(z)_{|z_+=0}
 \int dl_+d^2l_{\perp}
  e^{-i(\frac{l_+z_-}{2}-\pv{l}\cdot\pv{z})}
  A(l_+,\pv{l}).
\eeq
So we just need to consider the case $z_+=0$. For convenience the
above equation can be written in the form
\beq
T=\int dl_+d^2l_{\perp}\frac{1}{2}\int \frac{dz_-d^2z_{\perp}}{(2\pi)^3}
  e^{-i(\frac{l_+z_-}{2}-\pv{l}\cdot\pv{z})}M(z)_{|z_+=0}
  A(l_+,\pv{l}).
  \label{t2}
\eeq
Introduce Fourier transformation:
\beq
\Psi_B^\pm(l_+, l_{\bot})=
 \frac{1}{2} \int \frac{dz_-d^2z_{\perp}}{(2\pi)^3}
    e^{-i(\frac{l_+z_-}{2}-\pv{l}\cdot\pv{z})} \tilde\Psi^\pm_B(-{\pv{z}\,}^2,z_-/2)\
    ,
\eeq
and substitute eq.(\ref{m1}) into eq.(\ref{t2}), we can obtain
\beq
T=\int dl_+d^2l_{\perp} \frac{-if_B}{4}\left[
\frac{\dirac{p}_B+m_B}{2}\left\{
2\Psi^+_B(l_+,l_{\perp})+\frac{1}{2}\int \frac{dz_-d^2z_{\perp}}{(2\pi)^3}
    e^{-i(\frac{l_+z_-}{2}-\pv{l}\cdot\pv{z})} \right.\right. \nonumber\\
 \cdot \left.\left.
    \frac{\tilde\Psi^-_B(-{\pv{z}\,}^2,z_-/2)-\tilde\Psi^+_B(-{\pv{z}\,}^2,z_-/2)}{z_-/2}
    \dirac{z}\right\} \gamma_5
    \right]A(l_+,\pv{l}).
\eeq
By defining $\Delta(l_+, l_{\bot})\equiv\int_0^{l^+} dl
(\Psi^-_B(l, l_{\bot})-\Psi^+_B(l, l_{\bot}))$ and making the
integrations by part, we can finally get
\beq
T=\int dl_+d^2l_{\perp} \frac{-if_B}{4}\left[
\frac{\dirac{p}_B+m_B}{2}\left\{
2\Psi^+_B(l_+,l_{\perp})-\Delta(l_+, l_{\bot})\gamma^\mu
\frac{\partial}{\partial l^\mu} \right\}\gamma_5\right]
A(l_+,\pv{l}).
\label{t3}
\eeq
To deal with $\frac{\partial}{\partial l^\mu}$, we need to
re-express $l^\mu$ as
$$ l^\mu=\frac{l_+}{2}n_+^\mu+\frac{l_-}{2}n_-^\mu+l_{\perp}^\mu,
$$
therefore we can get
\beq
\frac{\partial}{\partial l^\mu}=n_-^\mu \frac{\partial}{\partial l_+}
    +n_+^\mu \frac{\partial}{\partial l_-}+\frac{\partial}{\partial
    l_{\perp}^\mu}.
\label{p1}
\eeq
Substitute eq.(\ref{p1}) into eq.(\ref{t3}) and do the integration
by part and drop the surface term, we get
\beq
T=\int dl_+d^2l_{\perp} \frac{-if_B}{4}\left[
\frac{\dirac{p}_B+m_B}{2}\left\{
\Psi^+_B(l_+,l_{\perp})\dirac{n}_+
+\Psi^-_B(l_+,l_{\perp})\dirac{n}_-
-\Delta(l_+, l_{\bot})\gamma^\mu
\frac{\partial}{\partial l_{\perp}^\mu}
\right\}\gamma_5\right]\nonumber\\
\cdot A(l_+,\pv{l}).~~~~~~~~
\eeq
So the projector for $B$ meson in the momentum space is
\beq
M_{\alpha\beta}=-\frac{if_B}{4}\left[
\frac{\dirac{p}_B+m_B}{2}\left\{
\Psi^+_B(l_+,l_{\perp})\dirac{n}_+
+\Psi^-_B(l_+,l_{\perp})\dirac{n}_-
-\Delta(l_+, l_{\bot})\gamma^\mu
\frac{\partial}{\partial l_{\perp}^\mu}
\right\}\gamma_5\right]_{\alpha\beta}.
\eeq


\begin{thebibliography}{99}

\bibitem{BBNS1} M. Beneke, G. Buchalla, M. Neubert and C.T. Sachrajda,
  \prl{83}{1999}{1914-1917}; \npb{591}{2000}{313-418};
  \npb{606}{2001}{245-321}.

\bibitem{pQCD} H. Li and H. Yu, \prl{74}{1995}{4388-4391};
  Y. Keum, H. Li and A.I. Sanda, \plb{504}{2001}{6};
    \prd{63} {2001}{054008};
  C. L\"u, K. Ukai and M. Yang,  \prd{63} {2001}{074009};
  C. L\"u and M. Yang, Eur. Phys. J. C23(2002)275.

\bibitem{Sachrajda} S. Descotes and  C.T. Sachrajda, \npb{625}
  {2002}{239-278}.

\bibitem{DHWY} D. Du, C. Huang, Z. Wei and M. Yang, \plb{520}{2001}{50-58}.

\bibitem{BLER} S.  Brodsky and G.  Lepage, \prl{43}{545}{1979};
   \plb{87}{1979}{359}; \prd{22}{1980}{2157};
  A. Efremov and A. Radyushkin, \plb{94}{1980}{245}.

\bibitem{Isgur} N. Isgur and C. Smith, \npb{317}{1989}{526-572}.

\bibitem{LiSterman}  H. Li and G. Sterman, \npb{381}{1992}{129-140}.

\bibitem{BBNS2} M. Beneke, G. Buchalla, M. Neubert and C.T. Sachrajda,
  \npb{606}{2001}{245-321}.

\bibitem{Liform} T. Kurimoto, H. Li and A.I. Sanda, \prd{65}{2002}{014007}.

\bibitem{Jakob} R. Jakob and P. Kroll, \plb{315}{1993}{463-470}.

\bibitem{Braun} E. Braun, P. Ball and V.M. Barun, \plb{417}{1998}{154-162};
  A. Khodjamirian, R. R$\ddot{\rm u}$ckl, S. Weinzierl and O. Yakovlev,
   \plb{410}{1997}{275-284}.

\bibitem{Brauntwist} V.M. Braun and I.E. Filyanov,
   \zpc{44}{1989}{157}; \zpc{48}{1990}{239}.

\bibitem{BenekeFeldmann} M. Beneke and Th. Feldmann,
   \npb{592}{2001}{3-34}.

\bibitem{GrozinNeubert} A.G. Grozin and M. Neubert, \prd{55}{1997}{272-290}.

\bibitem{Qiao}  H. Kawamura, J. Kodaira, C. Qiao and K. Tanaka,
   \plb{523}{2001}{111}; ibid., preprint: hep-ph/0112174.

\bibitem{LiSudakov} H. Li, \prd{52}{1995}{3958-3965}.

\bibitem{BottsSterman} J. Botts and G. Sterman, \npb{325}{1989}{62-100}.

\bibitem{Lithreshold} H. Li,  preprint: \hepph{0102013}.

\bibitem{Yao} R. Akhoury, G. Sterman and Y. Yao, \prd{50}{1994}{358}.

\bibitem{Brodsky} A. Szczepaniak, E.M. Henley and S. Brodsky,
     \plb{243}{1990}{287}.

\bibitem{BSW} M. Wirbel, B. Stech and M. Bauer, \zpc{29}{1985}{637}.

\bibitem{li1997} H. Li, \prd{55} {1997} {105}.

\end{thebibliography}
\end{document}